\documentclass[10pt]{iopart}

\usepackage{iopams}
\usepackage{epsfig}
\usepackage{mathrsfs}

\newcommand{\bz}{{\bf z}}
\newcommand{\bv}{{\bf v}}
\newcommand{\cD}{{\cal D}}
\newcommand{\grad}{{\boldsymbol{\nabla}}}

\begin{document}

\letter{Duality mapping and  unbinding
transitions of semiflexible and directed  polymers}
\author{Jan Kierfeld and Reinhard Lipowsky}
\address{MPI f\"ur Kolloid-- und Grenzfl\"achenforschung,\\
  D--14424 Potsdam, Germany}
\date{\today}

\begin{abstract} 
Directed polymers (strings)  and semiflexible polymers (filaments) 
are one-dimensional objects governed by tension and bending energy,
respectively. They  undergo unbinding transitions 
in the presence of a  short-range attractive potential.
 Using transfer matrix methods
we establish a duality mapping 
for filaments and strings between the 
restricted partition sums  in the absence and the presence of a
short-range attraction. 
This allows us to obtain exact 
results for the  critical  exponents related to the unbinding
transition, the 
 transition point and  transition order.  
\end{abstract}

\pacs{05.70.Fh, 64.60.Fr, 82.35.Gh, 87.15.Aa}

\section{Introduction}

Directed polymers (or ``strings'' in the following) 
are one-dimensional objects  governed by their tension which 
tends to minimize the contour length of the  polymer. Semiflexible 
polymers (or ``filaments'' in the following), on the other hand,  
are governed by their bending energy
which tends to straighten the polymer.  
In the presence of a  short-range attractive potential, these objects
 undergo unbinding or desorption transitions
which  represent a number of important 
critical phenomena \cite{FLN91,LL94}. The unbinding of strings describes 
wetting \cite{FLN91}, polymer adsorption \cite{adsorption}, 
pinning of flux-lines in type-II
superconductors \cite{flux}, or roughening of crystal surfaces
\cite{roughening}. 
The unbinding of filaments describes adsorption and 
bundling of many biopolymers (DNA, F-actin, microtubules) and 
 polyelectrolytes with large persistence lengths \cite{KL03}.

In this letter we use transfer matrix (TM) methods to derive 
a duality mapping for filaments and strings between
  the restricted partition sums 
 in the absence and in the presence of a 
short-range attractive potential. This  allows us to obtain 
the unbinding and desorption transition point, the order of the
transition,  and  a set of 
 scaling relations for the  critical exponents  of bound and unbound 
filaments and strings.

\section{Model}

We consider strings or filaments in $1+d_\perp$ dimensions 
which are oriented along the $x$-axis such that we can parameterize the 
contour by a $d_\perp$-dimensional  field $\bz(x)$ of displacements
perpendicular to the $x$-axis with $0<x<L$ where $L$ is the 
projected length of the string or filament.
The Hamiltonian for strings 
is given by the sum of the  tension energy
$\int_0^L dx  (\sigma/2) (\partial_x \bz)^2 $ 
with a string tension $\sigma$ and the potential energy 
$\int_0^L  V(\bz(x))$, where $V(\bz)$ contains an attractive 
potential well of
range $\ell_a$ which favours the configuration $\bz=0$. 
The Hamiltonian for filaments
is given by the sum of the  bending energy
$\int_0^L dx  ({\kappa}/{2}) (\partial^2_x\bz )^2$
and the potential energy $\int_0^L  V(\bz(x),\partial_x\bz)$.
 $\kappa$ is the bending rigidity of the filament and $L_p =
2\kappa/T$ the persistence length at  temperature $T$. 
The expression for the bending energy in the parameterization by 
the projected length is appropriate if {\em either}  the total length $L$ 
{\em or}  the longitudinal correlation length $\xi_\parallel$ to be
defined below are small compared to $L_p$.
In contrast to the string, the filament has 
 a well-defined tangent vector at each point, and therefore, 
also 
the external potential $V(\bz,\bv)$ can  depend on the tangent 
vector $\bv \equiv \partial_x \bz$.

Generic  potentials are of the form 
$V=V_r+V_a+V_p$ and  contain a 
hard-core potential $V_r$, a short-range attractive potential $V_a$, 
and eventually a long-range power-law potential $V_p$.
The hard core potential $V_{r}$ is given by  $V_{r}(\bz) = \infty$ for
$|\bz|<\ell_r$ and $V_{r}(\bz)=0$ otherwise.
   The  short-range 
  attractive potential $V_{a}$ has finite range $\ell_a$ 
  and a potential strength $W<0$, i.e., 
    $V_{a}(\bz) = W\Phi(\bv)$ 
   for $|\bz|<\ell_a$ ($\ell_a>\ell_r$) and 
   $V_a(\bz)=0$ otherwise. 
For strings  we can  only consider position-dependent potentials
and set  $\Phi(\bv)=1$. For filaments we include the  
dimensionless function $\Phi(\bv)$ modeling an additional
orientation-dependence of the attractive potential. 
The potential $V_a$ attains the asymptotic form 
$V_a(\bz) = G\ell_a^{-d_\perp}\Phi(\bv)\delta(\bz)$ in the limit of small
$\ell_a$ where $G \equiv W\pi^{d_\perp/2}/\Gamma(1+d_\perp/2)<0$. 
Finally, we can also include  attractive 
long-range power-law  potentials
$V_p(\bz)= w |\bz|^{-p}$ for $|\bz|>\ell_a$.
Our results apply
to potentials $V_p$ that decay sufficiently fast, i.e., potentials
$V_p$ with 
 $p\ge 2$ for strings 
and $p\ge 2/3$ for filaments \cite{L89}.

\section{Transfer matrix equations}

In order to simplify the notation, we introduce rescaled quantities 
measuring energies in units of the temperature $T$ and lengths in units of 
$T/2\sigma$ for strings and in units of the persistence length 
$L_p = 2\kappa/T$ for filaments.
In rescaled units the restricted partition sum for strings 
with fixed initial point 
 $\bz_0\equiv \bz(0)$ and end point  $\bz\equiv \bz(L)$
takes the form
\begin{equation}
\fl
  Z_L(\bz|\bz_0)  = 
    \int_{(\bz_0;0)}^{(\bz;L)}  \cD\bz(x)
   \exp{\left\{- \int_0^L dx \left[ \frac{1}{4}(\partial_x\bz )^2
       + V(\bz(x)) \right]
 \right\} }
~.
\label{Zsigma}
\end{equation}
In analogy with quantum mechanics,  this path-integral 
fulfils a Schr{\"o}dinger equation in imaginary time,
the partial differential TM equation given by 
\begin{equation}
 \partial_L Z_L = 
    \grad_{\bz}^2 Z_L  - V(\bz) Z_L
\label{TMsigma}
\end{equation}
with the boundary condition 
 $Z_0(\bz|\bz_0) = \delta(\bz-\bz_0)$ at $L=0$.
The  Laplace transform of the restricted partition sum with respect to
 $L$, 
$\tilde{Z}_s = \int_0^{\infty} dL e^{-sL} Z_L$,
 fulfils the  differential TM equation
\begin{equation}
 s \tilde{Z}_s = \grad_{\bz}^2 \tilde{Z}_s  - V(\bz) \tilde{Z}_s
   +\delta(\bz-\bz_0) 
\label{TMsigmaL}
\end{equation}
where the last term on the right hand side represents  the boundary 
condition at $L=0$.
For a sufficiently attractive potential, there exist bound states 
for which we make the Ansatz 
  $Z_L(\bz|\bz_0) \sim  \psi_E(\bz)\exp(-EL)$ 
where $E<0$ is the free energy
difference between the bound state and 
 the  free state (obtained   for $V=0$).  
The eigenfunction 
$\psi_E(\bz)$ for the energy level $E$ 
then  solves the stationary Schr{\"o}dinger equation (\ref{TMsigma})
\begin{equation}
- E \psi_E = \grad_{\bz}^2 \psi_E  - V(\bz) \psi_E
\label{TMsigmaE}
\end{equation}
with $E<0$ for a bound state. We  impose the normalization 
$\int_\bz \psi_E^2(\bz)=1$.
Then the solution 
 satisfying the proper boundary condition
is obtained by  summing over all energy levels $E_n$, 
$Z_L(\bz|\bz_0) = \sum_n \psi_{E_n}(\bz)\psi_{E_n}(\bz_0)e^{-E_nL}$,
where the 
ground state $E_0$  dominates the sum 
for lengths $L$ exceeding the correlation 
length $\xi_\parallel = 1/|E_0|$ (assuming that binding is 
weak such that the continuous scattering spectrum starts at $E_1=0$).

For  
filaments we can proceed similarly starting from 
the restricted partition sum in rescaled units, in which 
we additionally fix 
initial tangent  $\bv_0\equiv \partial_x\bz(0)$ and 
 end tangent 
 $\bv\equiv \partial_x\bz(L)$.
This partition function is given by
\begin{equation}
\fl
  Z_L(\bz,\bv|\bz_0,\bv_0)  = 
    \int_{(\bz_0,\bv_0;0)}^{(\bz,\bv;L)}  \cD\bz(x)
   \exp{\left\{- \int_0^L dx 
  \left[ \frac{1}{4}(\partial^2_x\bz )^2      
       + V(\bz(x),\partial_x \bz) \right]
 \right\} }
\label{Zkappa}
\end{equation}
and again  fulfils a
Schr{\"o}dinger-like differential TM equation \cite{GB89,MHL89} 
\begin{equation}
 \partial_L Z_L = - \bv \cdot \grad_{\bz}Z_L +
     \grad_{\bv}^2 Z_L  - V(\bz,\bv) Z_L
\label{TMkappa}
\end{equation}
with the boundary condition $Z_0(\bz,\bv|\bz_0,\bv_0) = 
\delta(\bz-\bz_0)\delta(\bv-\bv_0)$ at $L=0$.
As for strings we can consider the Laplace transform 
 which fulfils the 
  differential TM equation
\begin{equation}
 s \tilde{Z}_s = 
- \bv \cdot \grad_{\bz}\tilde{Z}_s +
   \grad_{\bv}^2 \tilde{Z}_s  - V(\bz,\bv) \tilde{Z}_s 
  +\delta(\bz-\bz_0)\delta(\bv-\bv_0)
\label{TMkappaL}
\end{equation}
where the last  term on the right hand side stems from the boundary
condition at $L=0$.
For sufficiently strong attractive potential, there exist bound states 
for which we make the Ansatz
  $Z_L(\bz,\bv|\bz_0,\bv_0) \sim 
\psi_E(\bz,\bv)\exp(-EL)$,  
where $E<0$ is the free energy
difference between bound and free state.
The eigenfunction 
$\psi_E(\bz,\bv)$ for the energy level $E$ 
then  solves the stationary version of 
the Schr{\"o}dinger-like equation (\ref{TMkappa})
\begin{equation}
- E \psi_E = - \bv \cdot \grad_{\bz}\psi_E + 
    \grad_{\bv}^2 \psi_E  -  V(\bz,\bv) \psi_E
\label{TMkappaE}
\end{equation}
with $E<0$ for a bound state. As for strings, 
we impose a normalization $\int_\bz\int_\bv
\psi_E(\bz,\bv)\psi_E(\bz,-\bv)=1$, and 
the solution 
 satisfying the proper boundary condition
is obtained by summing over all energy levels $E_n$.
For lengths $L$ exceeding  the correlation 
length $\xi_\parallel = 1/|E_0|$, the ground state dominates and 
 $Z_L(\bz,\bv|\bz_0,\bv_0) 
   \approx \psi_{E_0}(\bz,\bv)\psi_{E_0}(\bz_0,-\bv_0)e^{-E_0L}$.

\section{Scaling behaviour and exponents}

Strings and filaments differ in the scaling of free mean-square 
displacements, i.e.,  $\langle |\bz|^2 \rangle \sim L^{2\zeta}$
for $V=0$ where $\zeta$ is the roughness exponent. 
Strings show diffusive behaviour with $\zeta=1/2$, 
whereas filaments have $\zeta=3/2$. 
Tangent vector fluctuations 
scale as $\langle |\bv|^2 \rangle \sim L^{2(\zeta-1)}$
and show diffusive behaviour for filaments, 
whereas tangent vector fluctuations 
are finite and thus irrelevant for the scaling behaviour of strings. 
In the presence of a potential $V=V_r+V_a+V_p$, the scaling
behaviour of unbound {\em segments} of a string or filament is 
governed by the same roughness exponents (provided  $p\ge 2$ for strings 
and $p\ge 2/3$ for filaments \cite{L89}). 

For {\em unbound} strings and filaments, i.e., in the absence of a 
sufficiently strong attractive potential $V_a$, 
 this leads to  the scaling form 
\begin{equation}
 Z_L =  L^{-\chi_{u}} |\bz|^{\theta_{u}/2}
   \Omega_{u}\left(|\bz|L^{-\zeta},|\bv|L^{1-\zeta}\right) 
\label{Zscaling_unboundstringfilament}
\end{equation}
in the limit of small $|\bz_0|$ and $|\bv_0|$. For strings, the
tangent  $\bv$ is 
an irrelevant scaling variable.  We introduced exponents $\chi_{u}$
characterizing  the return probability and
$\theta_{u}$
 characterizing  the segment distribution at $\bz \approx 0$, 
and a shape function  $\Omega_{u}(y,u)$ 
 (with finite  $\Omega_{u}(0,0)$)
giving the shape of the 
 polymer segment distribution.

For strings and filaments {\em bound} by the attractive potential $V_a$,
the
longitudinal correlation length $\xi_{\parallel}=1/|E_0|$ 
gives the characteristic length of unbound segments and 
 enters the scaling behaviour, 
\begin{equation}
  Z_L =  \xi_{\parallel}^{-\chi_{b}}|\bz|^{\theta_{b}/2}  
   \Omega_{b}\left(z\xi_{\parallel}^{-\zeta},|\bv||\bz|^{(1-\zeta)/\zeta}
\right)   e^{L/\xi_{\parallel}}
 \label{Zscaling_boundstringfilament}
\end{equation}
with  analogous exponents $\chi_{b}$  and $\theta_{b}$, which differ
from the unbound case in general.

For a given potential, the two exponents $\chi$ and $\theta$ 
are not independent as can be seen by using the above scaling 
forms in the 
Chapman-Kolmogorov relations  $\int_\bz\int_\bv Z_L(\bz_1,\bv_1|\bz,\bv) 
Z_L(\bz,\bv|\bz_0,\bv_0) = 
Z_{2L}(\bz_1,\bv_1|\bz_0,\bv_0)$ for filaments
and $\int_\bz Z_L(\bz_1|\bz) Z_L(\bz|\bz_0) = 
Z_{2L}(\bz_1|\bz_0)$ for strings.
This  leads to scaling laws 
\begin{equation}
\fl
\chi = \max{\left(d_{\perp}/2+ \theta/2,0 \right)}
  ~~\mbox{(strings)},
~~~~
\chi = \max{\left(2d_{\perp}+ 3\theta/2,0
  \right)}~~\mbox{(filaments)},
 \label{scalinglaw}
\end{equation}
holding  both for $\chi_u$, $\theta_u$ and $\chi_b$, $\theta_b$. 
Exponents $\chi<0$ are not possible because they correspond to an 
unphysical {\em increase} of contacts as the length $\xi_{\parallel}$
of unbound segments increases. If $d_{\perp}/2+\theta/2<0$ for strings or $2d_{\perp}+ 3\theta/2<0$ for
filaments  a finite fraction of all polymer segments 
is bound at $\bz =0$ and the main contributions 
to the  $\bz$-integrals in the Chapman-Kolmogorov
relations come from small scales  $|\bz|\sim \ell_a$  leading to
$\chi=0$ in (\ref{scalinglaw}).

\section{Duality mapping}

Inspecting the Laplace transformed 
TM equation (\ref{TMsigmaL}) 
and the stationary TM equation
 (\ref{TMsigmaE}) for strings, we observe a formal similarity if we
identify $s=-E$: a
 short-range attractive 
 potential $V_a(\bz) \propto
-\delta(\bz-\bz_0)$  in the stationary TM equation
 (\ref{TMsigmaE})  plays the  role of 
 the initial condition in the Laplace transformed 
TM equation (\ref{TMsigmaL}) for a potential $V-V_a$, i.e., in the
absence  of the short-range attraction $V_a$. 
 A similar observation can be made for 
 the corresponding TM  equations (\ref{TMkappaL}) and
 (\ref{TMkappaE})
for filaments 
where a  short-range attractive 
  potential $V_a(\bz,\bv) \propto
 -\delta(\bz-\bz_0)\delta(\bv-\bv_0)$  in the stationary TM equation
  (\ref{TMkappaE})  plays the role of the initial condition in
the Laplace transformed 
TM equation (\ref{TMkappaL}) for a potential $V-V_a$. 
This is the main idea of the present paper and  will allow us to 
 establish a duality mapping between the stationary TM equation
 for {\em bound} states (characterized by the set of exponents
 $\theta_b$ and $\chi_b$) in 
a generic potential $V=V_r+V_a+V_p$ and the  Laplace transformed 
 TM equation for {\em unbound} states  (characterized by the set of exponents
 $\theta_u$ and $\chi_u$) in 
 a potential $V-V_a=V_r+V_p$ lacking the short-range
 attractive part.

A  string in a bound state $\psi_E^V(\bz)$
fulfils the  stationary TM equation
 (\ref{TMsigmaE}) for  a potential 
$V$ containing the short-range attraction
$V_a(\bz)= G \delta(\bz-\bz_0)$ where we consider the limit of small
$|\bz_0|$. We compare the stationary TM equation
 (\ref{TMsigmaE}) with 
the Laplace transformed 
TM equation (\ref{TMsigmaL}) for $\tilde{Z}_s^{V-V_a}(\bz|\bz_0)$
with   $s=-E$ and  for  a potential
$V-V_a$ without the short-range attraction.
If we rewrite 
$\delta(\bz-\bz_0)=
\delta(\bz-\bz_0)
\tilde{Z}_s^{V-V_a}(\bz|\bz_0)/\tilde{Z}_s^{V-V_a}(\bz_0|\bz_0)$
we find that both equations are equivalent and solutions have the same
normalization if the 
following two conditions are fulfilled:
\begin{eqnarray}
\fl
   \psi_E^{V}(\bz) &=&
   {\cal N}_E\tilde{Z}_{-E}^{V-V_a}(\bz|\bz_0) 
~~~~~~~~\mbox{with}~~{\cal N}_E^{-2}= \int_\bz
    [\tilde{Z}_{-E}^{V-V_a}(\bz|\bz_0)]^2
  \label{Zmapsigma}\\
\fl
  - G^{-1}&=&   \tilde{Z}_{-E}^{V-V_a}(\bz_0|\bz_0) = 
     \psi_E^{V}(\bz_0)/{\cal N}_E
~.
 \label{GEmapsigma}
 \end{eqnarray}
These two conditions define the duality 
mapping for strings 
between TM equations for potentials $V$ and $V-V_a$.

For filaments we proceed analogously for 
a bound state $\psi_E^V(\bz,\bv)$
which fulfils the  stationary TM equation
 (\ref{TMkappaE}) for  a potential 
$V$ containing the short-range attraction
$V_a(\bz,\bv)= G \delta(\bz-\bz_0)\delta(\bv-\bv_0)$,
where we consider the limit of small
$|\bz_0|$ and $|\bv_0|$. We compare the stationary TM equation
 (\ref{TMkappaE}) with the Laplace transformed 
TM equation (\ref{TMkappaL})  for 
$\tilde{Z}_s^{V-V_a}(\bz,\bv|\bz_0,\bv_0)$
with   $s=-E$ and  for a potential
$V-V_a$ without short-range attraction.
Following analogous steps as outlined for strings above, 
we find the following duality mapping for filaments, 
\begin{eqnarray}
\fl
  \psi_E^{V}(\bz,\bv) &=&
 {\cal N}_E\tilde{Z}_{-E}^{V-V_a}(\bz,\bv|\bz_0,\!\bv_0) 
\nonumber\\
\fl
  && ~~\mbox{with}~~{\cal N}_E^{-2}= \int_\bz\int_\bv
 \tilde{Z}_{-E}^{V-V_a}(\bz,\bv|\bz_0,\!\bv_0)
 \tilde{Z}_{-E}^{V-V_a}(\bz,-\bv|\bz_0,\!\bv_0)
 \label{Zmapkappa}\\
\fl
  - G^{-1}&=&   \tilde{Z}_{-E}^{V-V_a}(\bz_0,\!\bv_0|\bz_0,\!\bv_0) = 
     \psi_E^{V}(\bz_0,\bv_0)/{\cal N}_E
~,
 \label{GEmapkappa}
\end{eqnarray}
relating the TM equations for potentials $V$ and $V-V_a$.
This exact mapping  can be generalized to the more general class of 
potentials $V_a = G \Phi(\bv) \delta(\bz)$ if we use the additional 
assumption that 
$\tilde{Z}_{-E}^{V-V_a}(\bz_0,\bv_0|\bz_0, 0) \sim \delta(\bv_0)$ is 
a strongly localized function of $\bv_0$ in the limit 
$\bz_0\approx 0$. This assumption is justified  if the scaling 
function $\Omega_a(y,u)$ is exponentially decaying for $u\gg 1$ such
that  $\tilde{Z}_{-E}^{V-V_a}(\bz_0,\bv_0|\bz_0, 0) \approx 0$ for 
tangents $|\bv_0| \gg  |\bz_0|^{1/3}$. 
Then we can integrate both sides of (\ref{TMkappaL}) with a kernel
$\int_{\bv_0} \Phi(\bv_0) 
\tilde{Z}_{s}^{V-V_a}(\bz_0,\bv_0|\bz_0, 0) \ldots$,   which finally leads 
to a generalized duality mapping 
\begin{eqnarray}
\fl
  \psi_E^{V}(\bz,\bv) &=&
 {\cal N}_E\tilde{Z}_{-E}^{V-V_a}(\bz,\bv|\bz_0 ,0) 
\nonumber\\
\fl
 &&~~\mbox{with}~~{\cal N}_E^{-2}= \int_\bz\int_\bv
 \tilde{Z}_{-E}^{V-V_a}(\bz,\bv|\bz_0 ,0)
    \tilde{Z}_{-E}^{V-V_a}(\bz,-\bv|\bz_0 ,0)
 \label{Zmapkappa2}\\
\fl
  - G^{-1}&=&  \int_{\bv_0} 
   \Phi(\bv_0)   \tilde{Z}_{-E}^{V-V_a}(\bz_0,\bv_0|\bz_0, 0) = 
     \int_{\bv_0} 
   \Phi(\bv_0)  \psi_E^{V}(\bz_0,\bv_0)/{\cal N}_E
~,
 \label{GEmapkappa2}
\end{eqnarray}
which is valid in the limit  $\bz_0\approx 0$.

The validity of the duality mappings 
can be confirmed for a number of
potentials by direct TM calculations for strings \cite{L91} and 
filaments \cite{B93,KL03}.
The mappings allow us to obtain results for the full potential $V$ by
solving
 the Laplace transformed problem for the  simpler potential $V-V_a$ 
and give direct information on the partition sums
 $\psi_E^{V}$ and $\tilde{Z}_{s}^{V-V_a}$ and thus the segment distributions.
The duality  mappings  
 generalize exponent relations 
that have been found previously, as we will show in the 
following section. 
Furthermore,  
 relations  (\ref{GEmapsigma}), (\ref{GEmapkappa}) and  
(\ref{GEmapkappa2}) allow us 
 to determine the transition point, i.e., 
the critical potential strength $G_c$, 
and the  exponent $\nu_{\parallel}$ describing the 
 divergence of the correlation length close to the transition, 
$\xi_{\parallel} \propto
 |G-G_c|^{-\nu_{\parallel}}$.

\section{Exponent relations}

Without working out explicit solutions of the TM equations, 
we can use the duality mapping to derive various exact exponent
relations. 
To derive the exponent relation for $\chi_u$ and $\chi_b$ for strings 
we study 
the limit of small $|E|$ in  (\ref{Zmapsigma}).
The scaling form (\ref{Zscaling_unboundstringfilament}) for the 
unbound string determines the $s$-dependence of the singular part of 
$\tilde{Z}_s^{V-V_a}$  for small $s$ according to 
 $\tilde{Z}_{s,{\rm sing}}^{V-V_a} \sim  s^{\chi_{u}-1}$. 
For $\chi_{u}<1$ the singular part is the leading order contribution; for 
$\chi_{u}>1$ the leading order contribution is finite, 
 $\tilde{Z}_s^{V-V_a} \sim {\rm const}$. 
Using the Chapman-Kolmogorov relation, we find from (\ref{Zmapsigma}) 
the  singular 
behaviour ${\cal N}_E \sim |E|^{1-\chi_{u}/2}$ for $\chi_{u}<2$ 
for small $|E|$ 
and ${\cal N}_E \sim   {\rm const}$ for $\chi_{u}>2$. 
Furthermore, $\psi_E^{V} \sim |E|^{\chi_{b}/2}$ for small $|E|$ according to
the  scaling form (\ref{Zscaling_boundstringfilament}). 
Equating powers of $|E|$ in (\ref{Zmapsigma}) we arrive at the 
exponent  relation
\begin{equation}
   \chi_b = \left\{ 
     \begin{array}{ll}
     \max{\left(2- \chi_{u},0\right)}   &\mbox{for}~~ \chi_u>1 \\
     \chi_u                             &\mbox{for}~~ \chi_u<1
    \end{array}
   \right.
\label{exponents}
\end{equation}
for strings. For filaments, an  analogous analysis of  relation 
(\ref{Zmapkappa}) at small $|E|$  gives the {\em same}
 exponent relation (\ref{exponents}).
For strings, 
relation (\ref{exponents})
 agrees with direct calculations using the TM 
equations \cite{L91} and  also  applies 
in the presence of a long-range power-law potential $V_2 \sim w
|\bz|^{-2}$ ($p=2$), 
where the exponents $\chi$ depend continuously on $w$,  
as can be checked using the results of Ref.~\cite{LN88}.
Also for filaments, (\ref{exponents})  agrees with direct TM 
calculations for potentials $V=V_a$ and $V=V_r+V_a$ \cite{KL03,B93}. 
This exponent relation  has been formulated in Ref.~\cite{BLL00} 
  based on a  mapping between the 
 renormalization group equations for  strings and filaments of
 different dimensionality.  An equivalent exponent relation has been 
confirmed  numerically in Ref.~\cite{GB89}.

In order to derive the corresponding exponent relation 
for $\theta_u$ and $\theta_b$ for strings 
and filaments, we analyze the scaling behaviour  of the Laplace transform 
$\tilde{Z}_{s}^{V-V_a}$ of the unbound string or filament 
 for small $|\bz|$ in  (\ref{Zmapsigma}) and 
(\ref{Zmapkappa}), respectively.  Using the scaling form 
 (\ref{Zscaling_unboundstringfilament}) 
for the unbound string  or  filament we find
 $\tilde{Z}_{s}^{V-V_a} \sim |\bz|^{(1-\chi_u)/\zeta+\theta_u/2}$ for 
$\chi_u  >1$ and  $\tilde{Z}_{s}^{V-V_a} \sim |\bz|^{\theta_u/2}$ 
for $\chi_u <1$. 
According to the scaling form (\ref{Zscaling_boundstringfilament})
for the bound string or filament we have 
$\psi_E^{V} \sim |\bz|^{\theta_{b}/2}$ for small $|\bz|$.    
Equating powers of $|\bz|$ in (\ref{Zmapsigma}) or (\ref{Zmapkappa})
we arrive at the  exponent  relation
\begin{equation}
   \theta_b = \left\{ 
     \begin{array}{ll}
      \theta_u + 2(1-\chi_u)/\zeta  &\mbox{for}~~ \chi_u>1 \\
      \theta_u                      &\mbox{for}~~ \chi_u<1
    \end{array}
   \right.
\label{exponentstheta}
\end{equation}
which holds for strings with $\zeta=1/2$ and $\chi_u = d_{\perp}/2+
\theta_u/2$ {\em and} filaments with $\zeta=3/2$ and $\chi_u = 2d_{\perp}+
3\theta_u/2$, according to  the scaling laws (\ref{scalinglaw})
(note that $\chi_u>0$ for the unbound case).
For $\chi_b>0$,  the same exponent relation
can be obtained from a linear combination of (\ref{exponents}) and
the two relations which follow from (\ref{scalinglaw}) for the exponent
pairs $\chi_u, \theta_u$ and $\chi_{b}, \theta_{b}$, respectively.
Again, it can be checked that  relation (\ref{exponentstheta}) 
agrees with direct TM calculations both for strings  \cite{L91,LN88}
and for filaments \cite{KL03,B93,GB89,BLL00}.

Now we address the transition point, transition order, and the  
correlation length exponent  $\nu_\parallel$
by analyzing the dependence of the bound state energy $E$ on the
potential strength $G$ in relations (\ref{GEmapsigma}) and 
(\ref{GEmapkappa}). Setting $E=0$ on the right hand side we find the 
transition point $G_c$. 
As the singular part of $\tilde{Z}_s^{V-V_a}$  for small $s$
is $\tilde{Z}_{s,{\rm sing}}^{V-V_a} \sim  s^{\chi_{u}-1}$, we find $G_c=0$ 
for $\chi_u< 1$; thus, there is
no unbinding transition for $\chi_u< 1$  and 
strings and filaments are always in a bound state. 
Expanding around $E=0$ for $\chi_u>1$ gives 
$|G_c^{-1}-G^{-1}| \propto  |E|^{1/\nu_\parallel} =
\xi_\parallel^{-1/\nu_\parallel}$ with 
\begin{equation}
   1/\nu_\parallel= \min{(\chi_u-1,1)} ~~~\mbox{for}~~ \chi_u>1
~.
\label{nu}
\end{equation}
We also used that the linear order dominates the singular contribution 
to $\tilde{Z}_s^{V-V_a}$
for $\chi_u>2$ such that the transition becomes {\em first order} with
$\nu_\parallel=1$.  For $1<\chi_u<2$, we find $\nu_\parallel >1$ and a 
{\em continuous}  transition. 
The result  (\ref{nu}) agrees  with those of 
the necklace model \cite{F84}.
For filaments,  relation (\ref{nu}) 
can be generalized for  a class of tangent-dependent potentials 
 $V_a = G \Phi_\Delta(\bv) \delta(\bz)$, satisfying  a homogeneity 
relation $\Phi_\Delta(b\bv)= b^{-\Delta} \Phi_\Delta(\bv)$, 
which  has been considered also in Ref.~\cite{KL03}.
Performing the analogous  expansion in  (\ref{GEmapkappa2})
we find 
\begin{equation}
\fl 
   1/\nu_\parallel= \min{(\tilde{\chi}_u-1,1)} ~~~\mbox{for}~~ 
               \tilde{\chi}_u>1~,~~\mbox{where}~~ 
   \tilde{\chi}_u \equiv \chi_u - d_\perp(1-\Delta)/2
~.
\label{nu2}
\end{equation}
For this class of  potentials  there is
no transition for $\tilde{\chi}_u< 1$, a first order transition 
for $\tilde{\chi}_u> 2$  and a continuous transition for 
$1<\tilde{\chi}_u<2$. The result (\ref{nu}) 
is recovered for $\Delta=1$ and  $\Phi_1(\bv) = \delta(\bv)$.

The exponent relations (\ref{exponents}) and (\ref{nu}) or (\ref{nu2}),
together with the scaling law (\ref{scalinglaw}) allow us to calculate
all critical exponents of the unbinding problem 
 if only one exponent ($\chi_u$ or $\theta_u$) of the unbound string or
filament in the absence of the short-range attractive potential is 
known. These exponents are often known analytically, or can be easily 
obtained numerically. 
 For $V=0$, we have $\theta_u=\theta_0 =0$ for strings and
filaments. For $V=V_r$ and $d_\perp=1$,
 we can make use of another exponent relation, $\chi_u=\chi_r=1+\zeta$
 \cite{L95}, which is also valid for both strings and filaments.

\section{Conclusions}

In conclusion we derived  a duality mapping
between bound and unbound states of 
 one-dimensional strings and filaments. 
This mapping  allows us to determine the transition point and the order 
 of unbinding and desorption transitions of strings and filaments. 
We derived   exponent relations 
for the  return probability exponents $\chi$, the segment 
distribution exponents $\theta$  and the 
 correlation length exponent $\nu_\parallel$ from the mapping. 
These relations  allow us to determine 
{\it all critical exponents related to the 
 unbinding and desorption transitions of both filaments and strings
from a single   exponent characterizing the unbound
string or filament}.


\section*{References}



\begin{thebibliography}{99}

\bibitem{FLN91}
Forgacs G, Lipowsky R and Nieuwenhuizen T M 1991 {\it Phase
transitions and Critical Phenomena} Vol. 14, ed. by C. Domb and
J. Lebowitz (London, Orlando, FL: Academic Press) 

\bibitem{LL94}
L{\"a}ssig M and Lipowsky R 1994 
{\it Fundamental problems of statistical mechanics}, 
Vol. VIII (Amsterdam: Elsevier)



\bibitem{adsorption}
Eisenriegler E 1993  {\it Polymers near Surfaces} 
(Singapore: World Scientific)


\bibitem{flux}
Blatter G, Feigelman M V, Geshkenbein V B, 
  Larkin A I and Vinokur V M  1994 
  {\it Rev. Mod. Phys.} {\bf 66}  1125


\bibitem{roughening}
Krug J and Spohn H 1990  {\it Solids Far from Equilibrium: Growth,
Morphology and Defects}, ed. by C. Godr{\`e}che (Cambridge:
University Press) 

\bibitem{KL03}
Kierfeld J and Lipowsky R  2003 {\it  Europhys. Lett.} {\bf 62} 285  

\bibitem{L89}
Lipowsky R 1989 {\it Phys. Rev. Lett.} {\bf 62} 704

\bibitem{GB89}
Gompper G  and  Burkhardt T W 1989 {\it Phys. Rev. A} {\bf 40}  6124 

\bibitem{MHL89}
Maggs A C, Huse D A and Leibler S 1989  
 {\it Europhys. Lett.} {\bf 8} 615 

\bibitem{LN88}
Lipowsky R  and Nieuwenhuizen T 1988 
{\it  J. Phys. A: Math. Gen.} {\bf 21} L89

\bibitem{L91}
Lipowsky R 1991 {\it Europhys. Lett.} {\bf 15} 703 


\bibitem{B93}
Burkhardt T W 1993 {\it  J. Phys. A: Math. Gen.} {\bf 26} L1157 

\bibitem{BLL00}
Bundschuh R, L{\"a}ssig M and Lipowsky R 2000 
{\it Eur. Phys. J. E} {\bf 3} 295


\bibitem{F84}
Fisher M E 1984 {\it  J. Stat. Phys.} {\bf 34}  667, and references 
therein.

\bibitem{L95}
Lipowsky R 1995 {\it Z. Phys. B} {\bf 97} 193. 

\end{thebibliography}
\end{document}